\documentclass[conference]{IEEEtran}
\IEEEoverridecommandlockouts

\usepackage{cite}
\usepackage{amsmath,amssymb,amsfonts}
\usepackage{algorithmic}
\usepackage{graphicx}
\usepackage{textcomp}
\usepackage{xcolor}
\usepackage{bm}
\usepackage{url}
\usepackage{subcaption}
\newcommand{\hconj}{\mathsf{H}}
\usepackage{color}
\newcommand{\black}{\color{black}}

\def\BibTeX{{\rm B\kern-.05em{\sc i\kern-.025em b}\kern-.08em
    T\kern-.1667em\lower.7ex\hbox{E}\kern-.125emX}}
\begin{document}

\title{Spatio-Temporal Scheduling for Robust and Efficient Multi-Transmitter Wireless Power Transfer
\thanks{This paper is supported in part by JSPS KAKENHI Grant Number JP24K02933, JP26KJ1878 and MIYAKO-MIRAI Project of Tokyo Metropolitan University. We would like to thank Editage (www.editage.jp) for English language editing.}

}

\author{\IEEEauthorblockN{\IEEEauthorrefmark{2}Yuna Sawada, \IEEEauthorrefmark{3}Shino Shiraki, \IEEEauthorrefmark{2}$^{,}$\IEEEauthorrefmark{4}$^{,}$\IEEEauthorrefmark{5}Nobuyoshi Kikuma,\IEEEauthorrefmark{2}$^{,}$\IEEEauthorrefmark{5}Takahiro Matsuda,\\ \IEEEauthorrefmark{5}Takefumi Hiraguri, \IEEEauthorrefmark{5}$^{,}$\IEEEauthorrefmark{6}Kazuki Maruta, and \IEEEauthorrefmark{5}$^{,}$\IEEEauthorrefmark{7}Tomotaka Kimura}
\IEEEauthorblockA{\textit{\IEEEauthorrefmark{2}Graduate School of Systems Design, Tokyo Metropolitan University, Tokyo, Japan}\\
\textit{\IEEEauthorrefmark{3} Faculty of Innovative Information Science, Chiba Institute of Technology} \\
\textit{\IEEEauthorrefmark{4}Faculty of Engineering, Nagoya Institute of Technology, Nagoya, Japan} \\
\textit{\IEEEauthorrefmark{5}Faculty of Fundamental Engineering, Nippon Institute of Technology, Saitama, Japan}
\\
\textit{\IEEEauthorrefmark{6}Faculty of Engineering, Department of Electrical Engineering, Tokyo University of Science, Tokyo, Japan}\\
\textit{\IEEEauthorrefmark{7}Faculty of Science and Engineering, Doshisha University, Kyoto, Japan}\\
sawada-yuna@ed.tmu.ac.jp, 
takahiro.m@tmu.ac.jp, shiraki.shino@chibatech.ac.jp\\
kikuma@nitech.ac.jp,
hira@nit.ac.jp, maruta@rs.tus.ac.jp, 
tomkimur@mail.doshisha.ac.jp
}}

\maketitle

\begin{abstract}
In multi-user wireless power transfer (WPT), scheduling schemes that determine the allocation of transmission resources among receivers play a crucial role in improving power transfer efficiency.
Scheduling can be classified into time-division~(TD) and space-division~(SD) schemes, with the transmission order and direction designed to control when and to whom power is delivered.
TD-WPT can exploit the nonlinear characteristics of rectennas by concentrating power in time; however, a system relying on highly directional transmission from a single location reduces robustness under time-varying channel conditions.
This study investigated the effectiveness of spatio-temporal scheduling in coordinated multi-transmitter WPT systems.
By employing multiple transmitters, the proposed method is robust against channel variations in delivering power. Moreover, coordinated beamforming among transmitters exploits inter-cluster interference. Simulation results demonstrate the potential of the proposed scheme for robust and efficient power supply, even under shadowing conditions. 

\end{abstract}


\section{Introduction}

Wireless power transfer (WPT) has been attracting increasing attention as a promising technology for enabling sustainable Internet-of-Everything (IoE) networks, including batteryless sensor networks and Internet-of-Things (IoT) devices~\cite{Clerckx2021,Huang2019}.
In particular, far-field WPT, which utilizes radio frequencies as an energy resource, has received much attention because of its flexibility in device deployments and its capability to supply power over a wide area.
However, radio frequency signals experience significant propagation loss over distance, leading to a decrease in power transfer efficiency. 
To overcome this limitation, multi-antenna beamforming techniques, originally developed for wireless communications, have been applied to WPT~\cite{Shen2021,Zhou2018}.

In addition to improving power transfer efficiency, another important research direction focuses on \emph{scheduling}, including media access control (MAC) and system architecture design, for multi-user WPT to manage energy supply among multiple receivers~\cite{Bayat2022,Sawada2024,Sawada2025,Huang2019-2}.
Scheduling can be classified into two categories: time division~(TD) and space division~(SD).
In TD-WPT, an energy transmitter~(ET) periodically switches the beamforming weights to an assigned energy receiver~(ER). 
In SD-WPT, the ET simultaneously supplies power to multiple ERs by forming multiple beams with relaxed directivity.
Bayat and A\"{i}ssa~\cite{Bayat2022} demonstrated that TD scheduling outperforms SD scheduling under the max-min criterion owing to the nonlinear characteristics of the rectenna. 
This result shows that a concentrated power supply in a short duration is more efficient than using a wide beam pattern over an extended period.
In \cite{Sawada2024}, beneficial interference is considered, whereby signals intended for one receiver also supply energy to others, and the receiver grouping, which partially incorporates SD transmission, can further enhance the efficiency of pure TD scheduling. 
This result demonstrates the potential of interference in improving the power delivery efficiency of WPT.

In this paper, we propose a WPT system employing multiple transmitters to exploit beneficial inter-cluster interference. To realize this concept, we introduce \emph{spatio-temporal scheduling} for coordinated multi-transmitter WPT. In the proposed method, transmitters are spatially distributed, and each ET forms a \emph{cluster} that accommodates the nearby ERs. 
To leverage beneficial interference, all the ETs are connected to a central controller via a backhaul network and operate synchronously to optimize weight vectors and time slot assignment.  
The effectiveness of spatio-temporal scheduling in coordinated multi-transmitter WPT systems is further investigated in comparison with conventional collocated-transmitter multiple-input single-output~(MISO)-based configurations, focusing on robustness and energy efficiency across receivers.

The main contributions of this paper are summarized as follows:
\begin{itemize}
    \item We propose a new spatio-temporal scheduling architecture for multi-transmitter WPT systems. The proposed system enhances power transfer efficiency by combining three key techniques: receiver clustering, beamforming weight design for MISO transmission considering beneficial interference among transmitters, and time slot assignment.
 
    \item 
    It is natural to optimize scheduling from the perspective of fairness, and several previous studies have performed such optimization based on fairness~\cite{Bayat2022,Sawada2024,Sawada2025}.
    In~\cite{Kim2019}, the idea of $\alpha$-fairness is applied to waveform design in WPT systems. In this study, to achieve fair and efficient beamforming, we introduce an $\alpha$-fairness-based optimization framework to determine the MISO weight vectors. 
    By adjusting the fairness parameter $\alpha$, the proposed method can flexibly construct MISO beamforming according to different fairness criteria, such as max–min, proportional fairness, and sum-power maximization.
    In this paper, however, rather than directly evaluating fairness itself, we aim to evaluate the power transfer efficiency and robustness of the multi-transmitter WPT systems when optimized according to fairness criteria.

    \item The proposed framework was initially evaluated through two simulation scenarios. The results characterize the robustness of distributed ET deployment under shadowing and demonstrate the harvested-energy improvement achieved by inter-cluster-aware beamforming.
    
\end{itemize} 
The present evaluation focuses on the robustness of distributed ET deployment and the harvested-energy gain achieved by inter-cluster-aware beamforming, while a separate evaluation of optimized time-slot assignment is left for future work.

\section{System Model}
\label{sec:systemmodel}

\subsection{Cluster-based WPT}

\begin{figure}[t]
    \begin{subfigure}{\columnwidth}
        \centering
        \includegraphics[width=0.8\columnwidth]{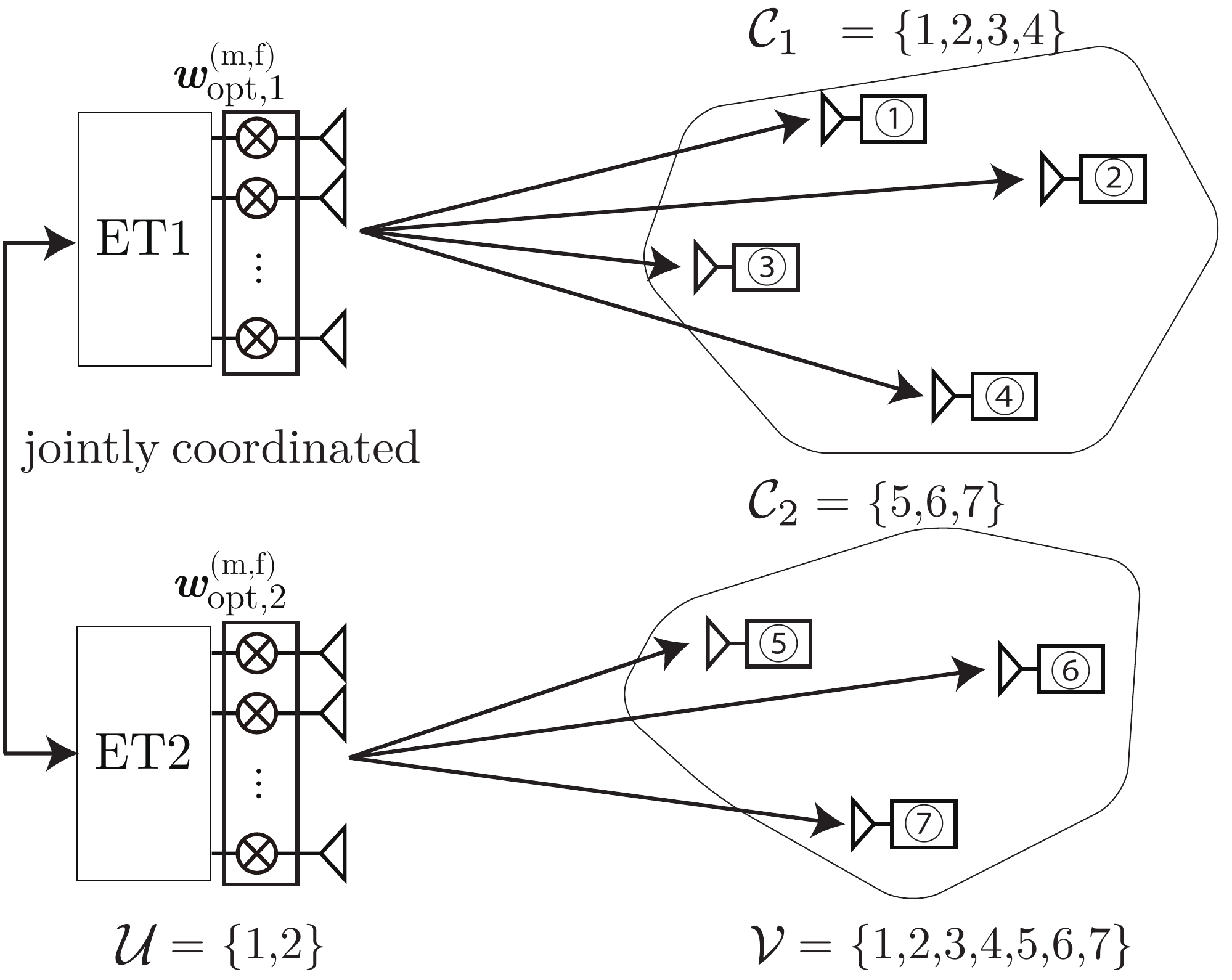}
        \caption{Cluster structure}
        \label{subfig:Clustersystem}
    \end{subfigure}\\[2ex]
    \begin{subfigure}{\columnwidth}
        \centering
        \includegraphics[width=0.85\columnwidth]{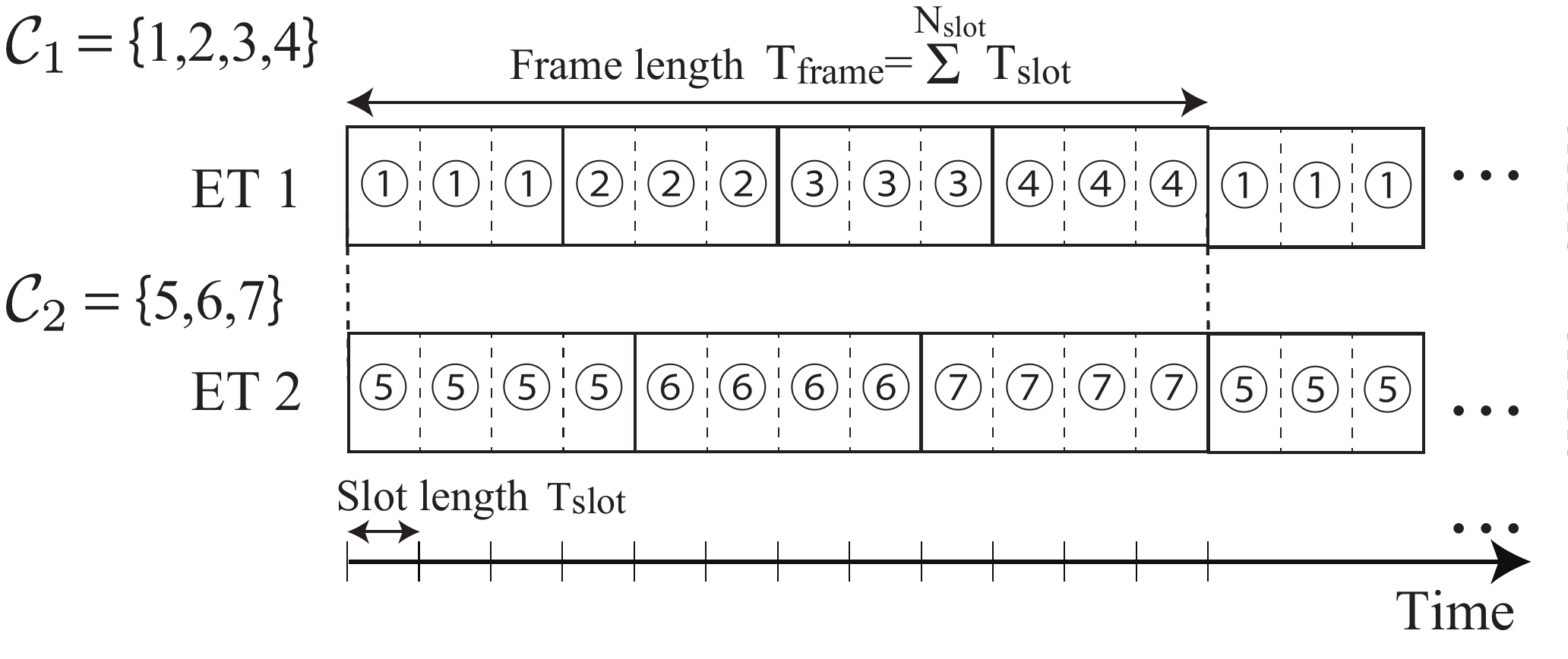}
        \caption{Structure of time slots}
        \label{subfig:TDsystem}
    \end{subfigure}
    \caption{System model~($N_\mathrm{ET} = 2,N_\mathrm{ER} = 7$).}
    \label{fig:system}
\end{figure}

Fig.~\ref{fig:system} shows the system model in this study. 
We consider a MISO-based WPT system consisting of $N_\mathrm{ET}$ ETs and $N_\mathrm{ER}$ ERs. Each ET is equipped with $L$ antenna elements and each ER is equipped with a single  rectenna. When an ER receives a radio signal transmitted by the ETs, the received RF signal is converted into DC power by the rectenna. The energy harvesting model is described in Section~\ref{subsec:energymodel}.

We define $\mathcal{U}=\{1,2,\ldots,N_\mathrm{ET}\}$ and $\mathcal{V}=\{1,2,\ldots,N_\mathrm{ER}\}$ as the sets of all ETs and ERs, respectively. 
ERs are divided into $N_\mathrm{ET}$ {\em{clusters}} $\mathcal{C}_n \subset \mathcal{V}$~($n = 1, 2, \ldots, N_{\mathrm{ET}})$ and each ER belongs to only one cluster. Let $\mathcal{C}= \{\mathcal{C}_n \subseteq \mathcal{V} \mid n = 1,2,\ldots,N_\mathrm{ET}\}$ denote the set of all clusters. $\mathcal{C}_n$~($n = 1, 2, \ldots, N_{\mathrm{ET}}$) satisfy the following conditions:
\begin{align*}
    \bigcup_{n=1}^{N_\mathrm{ET}}\mathcal{C}_n =\mathcal{V}&,\;\;
    \mathcal{C}_i \cap \mathcal{C}_j = \emptyset~(i\neq j)
\end{align*}
Fig.~\ref{subfig:Clustersystem} shows an example of $N_\mathrm{ET}=2$ and $N_\mathrm{ER}=7$. In this example, $N_\mathrm{ET}=2$ clusters $\mathcal{C}_1=\{1,2,3,4\}$ and $\mathcal{C}_2=\{5,6,7\}$ are formed.   

The channel vector between the $n$-th ET and the $k$-th ER is denoted by $\bm{h}_{n,k}^{(m,f)}=(h^{(m,f)}_{n,k,1}~h^{(m,f)}_{n,k,2}~\cdots~h^{(m,f)}_{n, k, L})^\top \in \mathbb{C}^{L}$, where $\top$ represents the transpose operator;
$h^{(m,f)}_{n,k,l}$~($l = 1, 2, \ldots, L$) denotes the channel coefficient between the $l$-th antenna element of the $n$-th ET and $\mathrm{ER}_k$ in the $m$-th slot of the $f$-th frame, which is referred to as the $(m, f)$ slot, hereafter.
In the proposed multi-transmitter WPT system, the channel vectors are obtained at the beginning of each frame, and the same channel vectors are used to optimize the weight vectors and slot assignments under the assumption that the channel vectors do not vary within the frame. Therefore, $\bm{h}_{n,k}^{(m,f)}$ can be expressed as $\bm{h}_{n,k}^{(f)}$. In the simulation experiments in Section~\ref{sec:simulation}, the impact of channel variations is evaluated using a time-varying wireless propagation model.

The structure of time slots is described in Section~\ref{subsec:TD-WPT}.
All ETs are connected to a central controller via a backhaul network, which enables them to share the channel vectors $\bm{h}_{n,k}^{(f)}~(n = 1,2,\ldots,N_\mathrm{ET}, k = 1,2,\ldots,N_\mathrm{ER})$ of all ET-ER links. In each slot, the $n$-th ET selects its target ER from $\mathcal{C}_n$, whereas the controller jointly optimizes the weight vectors of all ETs using the channels to all ERs selected in that slot.

This cluster-based structure forms the basis for coordinated energy transmission among multiple ETs, although unintended energy may still leak to other clusters as \emph{inter-cluster interference}.

\subsection{Frame structure and time slot assignment}
\label{subsec:TD-WPT}
Fig.~\ref{subfig:TDsystem} shows the time-slot structure assumed in this study.
The $n$-th ET switches the weight vectors for MISO beamforming periodically with a period of $T_{\mathrm{frame}}$, which denotes the length of the frames.
Each frame is further divided into slots with fixed length~$T_\mathrm{slot}$.
To define the smallest synchronized frame whose number of slots is divisible by the number of ERs in every cluster, we set
$N_\mathrm{slot}(\mathcal{C})=\mathrm{LCM}(|\mathcal{C}_1|,|\mathcal{C}_2|,\ldots,|\mathcal{C}_{N_\mathrm{ET}}|)$, that is, the least common multiple of the numbers of ERs across all clusters.
We assume that frames are synchronized across all the ETs, therefore, slots are also synchronized across all the ETs. 
Hereafter, $N_{\mathrm{slot}}(\mathcal{C})$ is simply denoted as $N_{\mathrm{slot}}$. 

The time slot assignment sequence~$\mathcal{S}^{(f)}$ in the $f$-th frame~($f = 1, 2, \ldots$) is defined as
\begin{align*}
    \mathcal{S}^{(f)} &=\left(S^{(1,f)}, S^{(2,f)}, \ldots,S^{(N_{\mathrm{slot}},f)}\right),\\
    \mathcal{S}^{(m,f)}&=\left\{s^{(m,f)}_n\in \mathcal{C}_n
    \middle| n=1,2,\ldots,N_\mathrm{ET}\right\},
\end{align*}
where $\mathcal{S}^{(m, f)}$ is the slot assignment in the $(m, f)$ slot~($m = 1, 2, \ldots, N_{\mathrm{slot}}$). 
In the proposed system, in each slot, each ET targets one ER in its cluster, and 
$s^{(m,f)}_n$ corresponds to the ER assigned to the $(m, f)$ slot.
In this study, $\mathcal{S}^{(f)}$ is optimized every frame as described in Section~\ref{subsec:Slot Assignments}.  
Note that in the $(m, f)$ slot, $s_n^{(m,f)}$ indicates the main target ER of the $n$-th ET for supplying energy. 
Namely, the $n$-th ET primarily directs its beam toward the ER $s_n^{(m,f)}$ to supply energy.

Suppose that $s_{n_0}^{(m, f)} = k$ , i.e., ER~($k\in\mathcal{C}_{n_0}$) is the target of ET $n_0$ in the $(m, f)$ slot.
Let $\bm{w}_n^{(m, f)}$ denote the weight vector of the $n$-th ET in the ($m, f$) slot.
With $\bm{w}_n^{(m, f)}$, the received signal at the $k$-th ER is expressed as
{\small
\begin{align}
    \nonumber
    y_k(\bm{w}^{(m,f)}) 
    =& \sum_{n = 1}^{N_{\mathrm{ET}}} (\bm{w}_n^{(m, f)})^\hconj\bm{h}_{n, k}^{(f)}e^{j2\pi f_{\mathrm{c}} t} \\
    \nonumber
    =& (\bm{w}_{n_0}^{(m, f)})^\hconj\bm{h}_{n_0, k}^{(f)}e^{j2\pi f_c t}
    \\
    \label{eq:rcvsignal}
    &+ \sum_{\substack{n^\prime = 1\\ n^\prime \neq n_0}}^{N_{\mathrm{ET}}} (\bm{w}_{n^\prime}^{(m, f)})^\hconj\bm{h}_{n^\prime, k}^{(f)}e^{j2\pi f_{\mathrm{c}} t},
\end{align}
}%
where $f_{\mathrm{c}}$ is the carrier frequency, and $\hconj$ represents the conjugate transpose operator. In~\eqref{eq:rcvsignal}, the second term corresponds to the inter-cluster interference. The received power of $y_k(\bm{w}^{(m,f)})$ is expressed as
\begin{align}
    \label{eq:rcvpower}
    \left|y_k(\bm{w}^{(m,f)})\right|^2 =& 
    \left|\sum_{n = 1}^{N_{\mathrm{ET}}}(\bm{w}_n^{(m, f)})^\hconj\bm{h}_{n, k}^{(f)}
    \right|^2\\
    \nonumber
    =&  (\bm{w}^{(m,f)})^\hconj \bm{H}^{(f)}_k  \bm{w}^{(m,f)}\\
    \nonumber
    \bm{w}^{(m,f)} =& \left((\bm{w}_1^{(m,f)})^\top~(\bm{w}_2^{(m,f)})^\top~\cdots~(\bm{w}_{N_{\mathrm{ET}}}^{(m,f)})^\top\right)^\top\\
    \nonumber
    \bm{H}^{(f)}_k =&
    \begin{pmatrix}
        \bm{H}^{(f)}_{1,1} & \bm{H}^{(f)}_{1,2} & \cdots & \bm{H}^{(f)}_{1,N_{\mathrm{ET}}}\\
        \bm{H}^{(f)}_{2,1} & \bm{H}^{(f)}_{2,2} & \cdots & \bm{H}^{(f)}_{2,N_{\mathrm{ET}}}\\
        \vdots & \vdots & \ddots & \vdots \\
        \bm{H}^{(f)}_{N_{\mathrm{ET},1}} & \bm{H}^{(f)}_{N_{\mathrm{ET},2}} & \cdots & \bm{H}^{(f)}_{N_{\mathrm{ET}},N_{\mathrm{ET}}}\\
    \end{pmatrix}\\
    \nonumber
    [\bm{H}^{(f)}_k]_{n_1, n_2} =& \bm{h}_{n_1, k}^{(f)}(\bm{h}_{n_2, k}^{(f)})^\hconj.
\end{align}
Based on \eqref{eq:rcvpower}, the received power may be enhanced or reduced according to $\bm{w}_n^{(m, f)}$~($n = 1, 2, \ldots, N_{\mathrm{ET}}$).  If the received power is enhanced, the inter-cluster interference is referred to as \emph{beneficial interference}, otherwise \emph{harmful interference}. 

In the ($m, f$) slot, ER$_k$ receives signals from all ETs, and the contributions from nonassociated $n^\prime$-th ETs~($n\neq n^\prime$) constitute inter-cluster interference. Therefore, to make the inter-cluster interference beneficial, the weight vectors $\bm{w}_{n}^{(m, f)}$~($n = 1, 2, \ldots, N_{\mathrm{ET}}$) should be optimized according to $\mathcal{S}^{(m, f)}$. 
The set of the optimized weight vectors is expressed as 
$\bm{w}_{\mathrm{opt}}^{(m,f)} = ((\bm{w}_{\mathrm{opt}, 1}^{(m,f)})^\top~(\bm{w}_{\mathrm{opt}, 2}^{(m,f)})^\top~\cdots~(\bm{w}_{\mathrm{opt}, N_\mathrm{ET}}^{(m,f)})^\top)^\top \in \mathbb{C}^{N_\mathrm{ET}L\times 1}$.
The weight vector $\bm{w}_{\mathrm{opt},n}^{(m,f)}$ for the $n$-th ET is obtained using the channel vectors $\bm{h}_{n,k}^{(f)}$~($\forall k \in \mathcal{S}^{(m, f)}$) of the ER assigned to the time slot~${s}^{(m,f)}_n$,  setting $\|\bm{w}_{\mathrm{opt},n}^{(m, f)}\|^2 \leq P/N_\mathrm{ET}$, where $\|\bm{w}_{\mathrm{opt},n}^{(m, f)}\|$ and $P$ denote the $\ell_2$-norm of $\bm{w}_{\mathrm{opt},n}^{(m, f)}$ and the total transmit-power, respectively. The weight vector optimization scheme is explained in Section~\ref{subsec:optw}. 

\begin{figure}
    \centering
    \includegraphics[width=0.6\columnwidth]{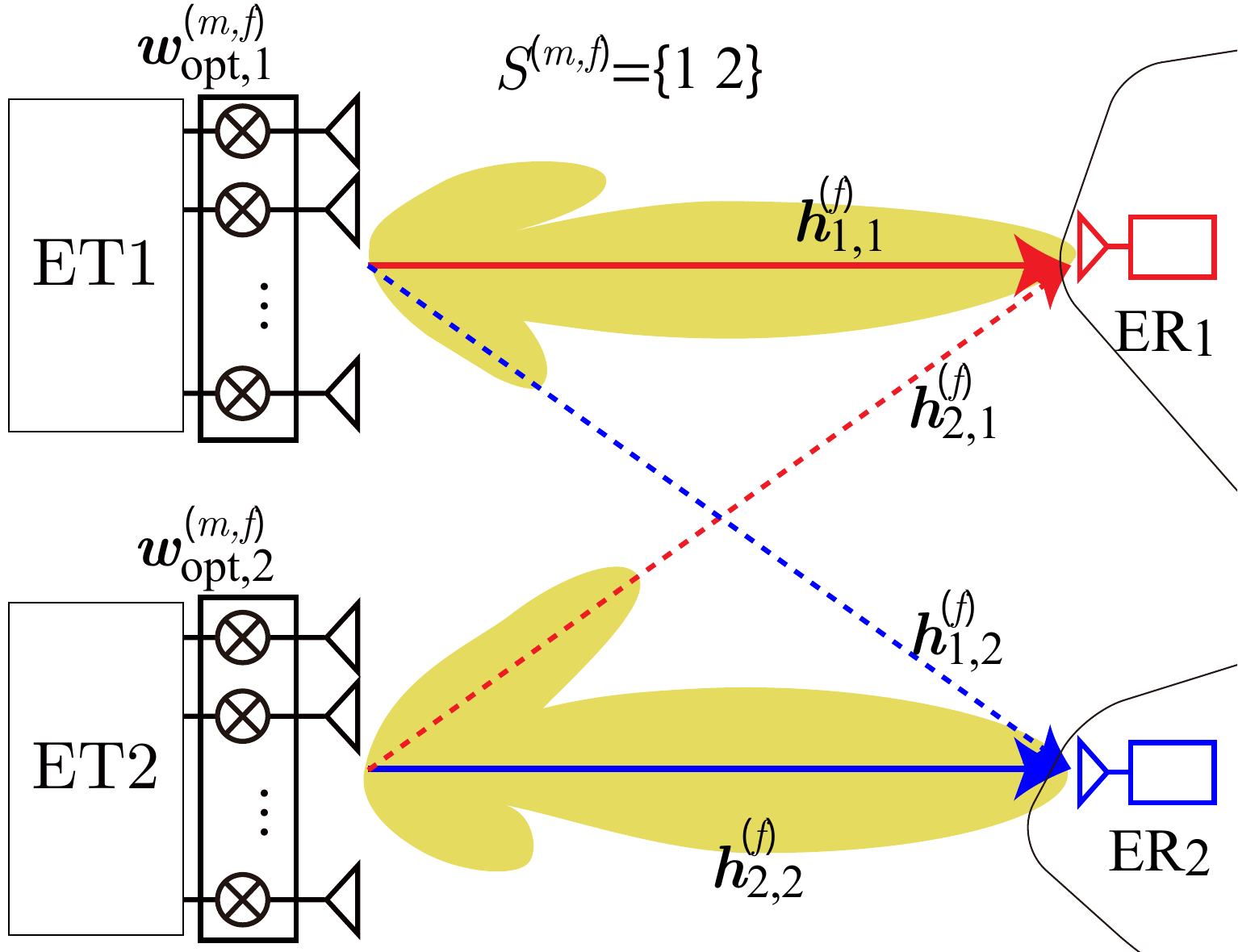}
    \caption{Weight and channel vectors in a coordinated multi-transmitter WPT system.}
    \label{fig:optweight}
\end{figure}

The relationship between weight and channel vectors is displayed in Fig.~\ref{fig:optweight} for $N_{\mathrm{ET}} = 2$, $N_{\mathrm{ER}} = 2$. Because there are only two one-element clusters $\mathcal{C}_1 = \{1\}$ and $\mathcal{C}_2 = \{2\}$, $N_{\mathrm{slot}}$ is set to $N_{\mathrm{slot}} = 1$. $\mathcal{S}^{(1, f)}$ is set to $\mathcal{S}^{(1, f)} = \{1, 2\}$, and $\bm{w}_{\mathrm{opt}}^{(m,f)} = ((\bm{w}_{\mathrm{opt},1}^{(1,f)})^\top~(\bm{w}_{\mathrm{opt},2}^{(1,f)})^\top)^\top$ is optimized using four channel vectors~$\bm{h}_{1,1}^{(f)},\bm{h}_{2,1}^{(f)},\bm{h}_{1,2}^{(f)},\bm{h}_{2,2}^{(f)}$.
The channel $\bm{h}_{1,1}^{(f)}$ and $\bm{h}_{2,2}^{(f)}$~(solid lines in Fig.~\ref{fig:optweight}) supply power to the receivers assigned to their own clusters, whereas the  channel $\bm{h}_{2,1}^{(f)},\bm{h}_{1,2}^{(f)}$~(dotted lines in Fig.~\ref{fig:optweight}) contribute to inter-cluster interference.

\subsection{Energy Harvesting Model}
\label{subsec:energymodel}
\begin{table}[b]
    \caption{Parameters for harvested energy $q_k(\bm{w}_k^{(m,f)})$~\cite{Bayat2022}}
    \vspace{-0.4em}
    \label{table:1}
    \begin{center}
    \begin{tabular}[t]{c||c}
    parameter & value \\
    \hline 
    $R_{\mathrm{L}}$ & 1 [$k\Omega$]\\
    $I_{\mathrm{s}}$ & 5[$\mu$A]\\
    $\eta$ & 1.05\\
    $v_{\mathrm{T}}$ & 26 [mV]\\
    \hline 
    \end{tabular}
        \begin{tabular}[t]{c||c}
    parameter & value \\
    \hline 
    $\lambda$ & 1\\
    $R_{\mathrm{ant}}$ & 50 [$\Omega$]\\
    \hline 
    \end{tabular}
    \end{center}
\end{table}

Let us define $q_k\bigl(\bm{w}_{\mathrm{opt}}^{(m,f)}\bigr)$
and $i_{\mathrm{out},k}\bigl(\bm{w}_{\mathrm{opt}}^{(m,f)}\bigr)$ as the rectenna output power and current by $\mathrm{ER}_k$ using a set $\bm{w}_{\mathrm{opt}}^{(m,f)}$ of the optimized weight vectors, respectively. 
According to the model in~\cite{Bayat2022}, $q_k(\bm{w}_{\mathrm{opt}}^{(m,f)})$ 
can be expressed as follows:
{\small
\begin{align}
    \label{eqn:q}
    &q_k\bigl(\bm{w}_{\mathrm{opt}}^{(m,f)}\bigr) = R_\mathrm{L}i^2_{\mathrm{out},k}\bigr(\bm{w}_{\mathrm{opt}}^{(m,f)}\bigl)\\
    \nonumber
    &i_{\mathrm{out},k}\bigl(\bm{w}_{\mathrm{opt}}^{(m,f)}\bigr) \\ &= \frac{1}{\mu} W_0\;\biggl(\mu I_s I_0\;\biggl(\frac{\bigl|y_k(\bm{w}_{\mathrm{opt}}^{(m,f)})\bigr| \; \sqrt{2 \lambda R_\mathrm{ant}}}{\eta v_\mathrm{T}} \biggr)\;e^{\mu I_s}\biggr) - I_s,
    \label{eq:signalsum}
\end{align}}%
where $\mu=R_{\mathrm{L}}/(\eta v_{\mathrm{T}})$, and $W_0(\cdot)$ represents the principal branch of the Lambert W function \cite{Lambert}; $I_{0}(\cdot)$ is the $0$-th order modified Bessel function of the first kind.
Table \ref{table:1} summarizes the parameters adopted in this study, which are the same values as in \cite{Bayat2022}.
$q_k\bigl(\bm{w}_{\mathrm{opt}}^{(m,f)}\bigr)$ is maximized if $|y_k(\bm{w}_{\mathrm{opt}}^{(m,f)})|$ is maximized because $W_0(x)$ and $I_0(x)$ are increasing functions for $x \geq 0$.

The harvested energy in the $f$-th frame and $(m, f)$ slot are represented respectively as follows:
\begin{align}
\nonumber
    Q_{k}(\mathcal{S}^{(f)}) &= \sum_{m=1}^{N_\mathrm{slot}} Q_{k}(\mathcal{S}^{(m,f)})\\
    \label{eq:Qkmf}
    Q_{k}(\mathcal{S}^{(m,f)}) &= T_\mathrm{slot} \;q_k\bigl(\bm{w}_{\mathrm{opt}}^{(m,f)}\bigr)
\end{align}

\section{Spatio-Temporal Scheduling in \\Coordinated Multi-Transmitter WPT}
\label{sec:STscheduling}
This section describes the proposed spatio-temporal scheduling method that uses multiple coordinated transmitters. This method consists of ER clustering, weight vector optimization, and time slot assignments. These components are explained in the following subsections. 

By using the set $\mathcal{C}$ of clusters obtained from the ER clustering, $\bm{w}^{(m,f)}$ and $\mathcal{S}^{(m,f)}$ are jointly optimized over each frame. The optimization problem can be formulated as follows:
\begin{align*}
    \max_{\{\bm{w}^{(m,f)}, \mathcal{S}^{(m, f)}\}_{m=1}^{N_\mathrm{slot}}}\min_{k \in \mathcal{V}} &\quad Q_k(\mathcal{S}^{(f)})\\
    \mbox{subject to} &\quad \|\bm{w}_{n}^{(m,f)}\|^2 = {\black P/N_\mathrm{ET}} \quad \forall n, m\\
    &\quad s_n^{(m, f)} \in \mathcal{C}_n \quad \forall n \in \mathcal{U}
\end{align*}
In the optimization problem, the max-min criterion is adopted to supply fair energy to all the ERs within each frame. 
Because this problem is a mixed-integer nonlinear optimization problem, we decompose it into beamforming optimization and frame-level slot-assignment selection, as explained in the following subsections.

\subsection{ER Clustering}
To increase signal strength, each ER should be assigned to the nearest ET cluster.
However, if the spatial distribution of ERs is biased, a simple distance-based clustering method may cause some clusters to include more ERs than others, resulting in insufficient power supply to the ERs belonging to the larger clusters.
Therefore, in the proposed method, the ERs are clustered according to the following round-robin-based clustering algorithm.
\begin{enumerate}
    \item Set $n := 1$ and $\tilde{\mathcal{V}} := \mathcal{V}$.
    \item The $k$-th ER with the minimum distance to the $n$-th ET is assigned to $\mathcal{C}_n$, and the set is updated as $\tilde{\mathcal{V}} := \tilde{\mathcal{V}} \setminus \{k\}$.
    \item If $\tilde{\mathcal{V}} = \emptyset$, terminate the algorithm; otherwise, set $n := n + 1$. If $n > N_{\mathrm{ET}}$, set $n := 1$, and return to Step 2.
\end{enumerate}

\subsection{Weight Vector Optimization}
\label{subsec:optw}

From~\eqref{eq:Qkmf}, for a given $\mathcal{S}^{(m,f)}$, the weight vectors $\bm{w}_n^{(m,f)}$~($n = 1, 2, \ldots, N_{\mathrm{ET}}$) can be optimized by solving 
\begin{align*}
    \max_{\bm{w}^{(m,f)}} &\quad \sum_{k \in \mathcal{S}^{(m, f)}} q_k(\bm{w}^{(m,f)})\\
    \mbox{subject to} &\quad \|\bm{w}^{(m,f)}_{n}\|^2 = {\black P/N_\mathrm{ET}} \quad \forall n \in \mathcal{U}.
\end{align*}
To obtain a computationally convenient beamforming objective, we replace the harvested-energy objective with a received-power-based surrogate.
The weight vector optimization problem is formulated as follows:
\begin{align*}
    \max_{\bm{w}^{(m,f)}}\quad & 
    U_{\alpha}(\bm{w}^{(m,f)}; \mathcal{S}^{(m,f)}) \\
    \nonumber
    \text{subject to} & \quad \|\bm{w}^{(m,f)}_{n}\|^2 ={P/\black N_\mathrm{ET}} \quad \forall n \in \mathcal{U}
\end{align*}
The objective function $U_\alpha(\cdot)$ is based on the $\alpha$-utility function for \emph{$\alpha$-fairness}~\cite{Chen2023}. $U_{\alpha}(\bm{w}^{(m,f)})$ is given by
\begin{align*}
    U_{\alpha}(\bm{w}^{(m,f)}; \mathcal{S}^{(m,f)}) = 
    \begin{cases}
        \displaystyle \dfrac{1}{1 - \alpha} \sum_{k \in \mathcal{S}^{(m, f)}}
        \left|y_k(\bm{w}^{(m,f)})\right|^{2(1 - \alpha)} & \\
        & \hspace{-6em} \alpha \geq 0, \alpha \neq 1\\
        \displaystyle \sum_{k \in \mathcal{S}^{(m, f)}} \log\left(\left|y_k(\bm{w}^{(m,f)})\right|^2\right)\\
        & \hspace{-3em} \alpha = 1\\
    \end{cases}
\end{align*}
The parameter $\alpha$ can be interpreted as quantifying the fairness/efficiency trade-off~\cite{Chen2023}.
For $\alpha \rightarrow \infty$, the optimization problem can be interpreted as a max-min optimization problem, where the weight vectors are optimized so as to maximize the minimum received power among the ERs in $\mathcal{S}^{(m, f)}$. 
For $\alpha = 0$, $U_{\alpha}(\bm{w}^{(m,f)}; \mathcal{S}^{(m, f)})$ is expressed as
\begin{align*}
    U_0(\bm{w}^{(m,f)}; \mathcal{S}^{(m, f)}) =& \sum_{k \in \mathcal{S}^{(m, f)}}\left|y_k(\bm{w}^{(m,f)})\right|^2
\end{align*}
Therefore, the weight vectors are optimized to maximize the sum of the received power at all the ERs in $\mathcal{S}^{(m, f)}$. 

For $\alpha = 1$, the optimization problem can be interpreted as a {\em proportional-fair} optimization problem to maximize the sum of the logarithms of the received powers. 
When $\alpha = 1$, $U_{\alpha}(\bm{w}^{(m,f)};\mathcal{S}^{(m,f)})$ is redefined as
\begin{align*}
    U_1(\bm{w}^{(m,f)}; \mathcal{S}^{(m,f)}) =
    \sum_{k\in\mathcal{S}^{(m,f)}} 
    \log\left(
    \left|y_k(\bm{w}^{(m,f)})\right|^2 + \epsilon
    \right),
\end{align*}
where $\epsilon > 0$ is a small regularization constant to avoid evaluating $\log 0$. 
Based on preliminary evaluations with several values of $\alpha$, we set $\alpha = 1$ in this paper. The proportional-fair utility $U_1(\bm{w}^{(m,f)}; \mathcal{S}^{(m,f)})$ effectively suppresses excessive power concentration and achieves a more balanced received power distribution among the ERs. A systematic evaluation of the various values of $\alpha$ is left for future work.

\subsection{Time Slot Assignments}
\label{subsec:Slot Assignments}
Let $\tilde{\mathcal{S}}_{i}^{(m,f)}$~($i = 1, 2, \ldots, N_{\mathrm{S}}$) represent the candidate slot assignments for the $(m,f)$ slot, where $N_{\mathrm{S}}$ denotes the number of candidate slot assignments in each slot. $N_{\mathrm{S}}$ depends on the clusters $\mathcal{C}_n$~($n = 1, 2, \ldots, N$) and is obtained by
\begin{align*}
    N_{\mathrm{S}} = \prod_{n = 1}^{N_{\mathrm{ET}}}|\mathcal{C}_n|
\end{align*}
Suppose that $N_{\mathrm{ET}} = 2$, $N_{\mathrm{ER}} = 4$, $\mathcal{C}_1 = \{1, 2\}$ and $\mathcal{C}_2 = \{3, 4\}$ with $N_{\mathrm{slot}} = 2$.
The $N_{\mathrm{S}} = 4$ candidate slot assignments are then given by
\begin{align*}
    \tilde{\mathcal{S}}_1^{(m,f)} = \{1, 3\}, \tilde{\mathcal{S}}_2^{(m,f)} = \{1, 4\}, \\
    \tilde{\mathcal{S}}_3^{(m,f)} = \{2, 3\}, \tilde{\mathcal{S}}_4^{(m,f)} = \{2, 4\}.
\end{align*}
For a given $\tilde{\mathcal{S}}_i^{(m,f)}$, the optimum weight vector $\bm{w}_{\mathrm{opt}}^{(m,f)}$ can be obtained by solving the optimization problem explained in Section~\ref{subsec:optw}.

The optimum slot assignment is obtained by solving a max-min problem, namely, selecting the sequence of time slot assignments that maximizes the supplied energy to the ER with minimum harvested energy per frame.  
An index sequence $I$ of the candidate slot assignments in the $f$-th frame is defined as
\begin{align*}
   I =& (i_1,i_2,\ldots,i_{N_\mathrm{slot}})
   , \;\;i_m \in \{1, 2, \ldots, N_{\mathrm{S}}\}
\end{align*}
and $\mathcal{I}=\{1,2,\ldots,N_S\}^{N_\mathrm{slot}}$ is defined as the set of all index sequences.
We also define $\tilde{Q}(I)$ as
\begin{align*}
    \tilde{Q}(I) =
    \min_{k \in \mathcal{V}}\left\{\sum_{m=1}^{N_\mathrm{slot}}
    Q_k(\tilde{\mathcal{S}}_{i_m}^{(m,f)})\middle| \forall k \in \mathcal{V}
    \right\}.
\end{align*}
The optimum slot assignment is obtained by solving the following problem:
\begin{align*}
    I^* &= \arg\max_{I\in \mathcal{I}} 
    \tilde{Q}(I) \\
    \mathcal{S}^{(f)}_\mathrm{opt} &= \left(\tilde{\mathcal{S}}_{i_1^*}^{(1,f)},\tilde{\mathcal{S}}_{i_2^*}^{(2,f)},\ldots,\tilde{\mathcal{S}}_{i_{N_\mathrm{slot}}^*}^{(N_\mathrm{slot},f)} \right)
\end{align*}
At the beginning of each frame, the ET first obtains the optimized weight vectors for all candidate $N_S$ slot assignments and then determines the frame-level sequence of slot assignments by solving the above max-min combinatorial optimization problem.
This procedure requires weight-vector optimization for all candidate slot assignments, resulting in high computational complexity as the number of ERs increases.
We will consider an efficient algorithm to solve this problem in future research. 
A genetic algorithm~(GA) that encodes each assignment-index sequence as an individual is a promising approach for reducing the search complexity.

\section{Performance Evaluation}
\label{sec:simulation}

\subsection{Simulation Environment}
\label{sec:simulationenvironment}
    \begin{figure}
        \centering
        \includegraphics[width=0.5\linewidth]{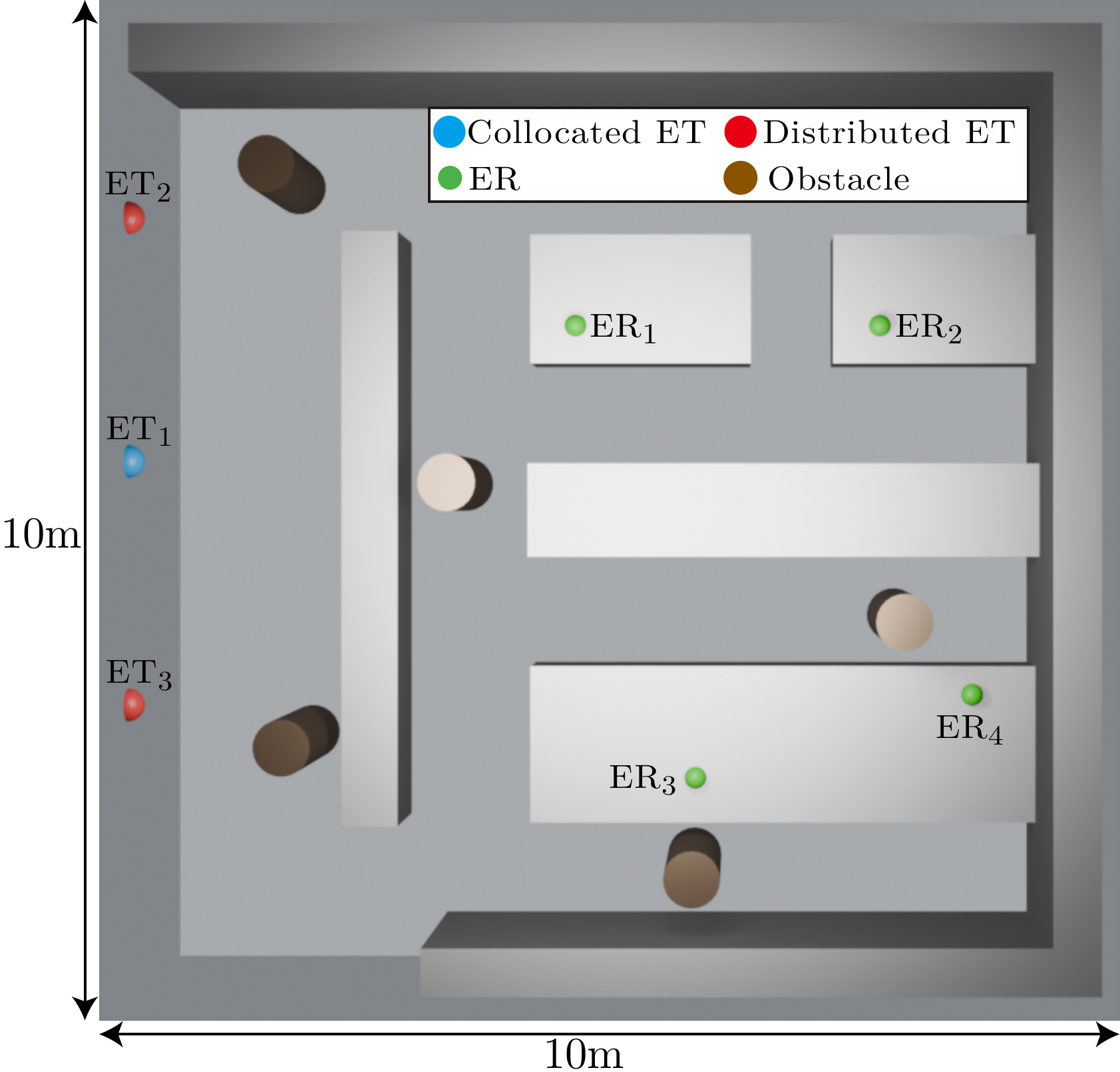}
        \caption{Simulation environment.}
        \label{fig:simenvironment}
    \end{figure}

As shown in Fig.~\ref{fig:simenvironment}, an office environment of $10 \mathrm{m} \times 10 \mathrm{m}$ with desks and shelves was considered for two simulation scenarios. In the first scenario, we evaluated the proposed system with $N_{\mathrm{ET}} = 2~($ET$_2$,ET$_3$) and $N_{\mathrm{ER}} = 2$, representing ``distributed'' deployment against a WPT system with $N_{\mathrm{ET}} = 1~($ET$_1)$ and $N_{\mathrm{ER}} = 2$, representing ``collocated'' deployment. 
In this scenario, ERs were placed at ER$_1$ and ER{\black $_3$} in Fig.~\ref{fig:simenvironment}, and the ETs are placed at a height of $2.0$~m. 
In the distributed system, ET$_2$ and ET$_3$ are each equipped with a $32$-element planar array~($4$ rows and $8$ columns), whereas the collocated system uses a single $64$-element planar array~($8$ rows and $8$ columns) at ET$_1$.
Both systems therefore used the same total number of antenna elements and the same total transmit power of $P=32~\mathrm{W}$. In the distributed system, ET$_2$ and ET$_3$ simultaneously targeted ER$_1$ and ER$_3$, respectively, whereas ET$_1$ alternately targeted ER$_1$ and ER$_3$ in the collocated system. Each ER was targeted in every slot in the distributed system and in one of the two slots in the collocated system.

In the second scenario, we compared two beamforming methods: an \emph{inter-cluster-aware} method and a \emph{cluster-wise} method. Both methods used the same distributed configuration  with $N_\mathrm{ET}=2,N_\mathrm{ER}=4$. ER$_1$–ER$_4$ were placed at the locations shown in Fig.~\ref{fig:simenvironment}, and each ET was equipped with a 32-element planar array. In the inter-cluster-aware method, the weight vectors of all ETs were jointly optimized using the channels to all ERs selected in the current slot, as described in Section~\ref{sec:STscheduling}. The cluster-wise method used the same slot assignment as the inter-cluster-aware method; however, each ET independently applied maximum-ratio transmission (MRT) toward the ER selected from its own cluster, without considering its effect on ERs in the other clusters.

In both scenarios, the channel vectors $\bm{h}_{n, k}^{(f)}$ were generated at $5.7$~GHz using the Sionna RT ray-tracing simulator~\cite{SionnaRT}. The static objects were modeled as metal, whereas the floor and the walls were modeled as concrete. As shown in Fig.~\ref{fig:simenvironment}, five moving obstacles followed a random waypoint model to induce time-varying LoS blockage and multipath variations. 
The simulation experiments with a slot time of $T_\mathrm{slot} = 0.1~\mathrm{s}$ were conducted for a duration of $100~\mathrm{s}$.
\subsection{Simulation Results}
\subsubsection{Effect of Distributed ET deployment}
\begin{figure}[t]
    \centering
    \begin{subfigure}[t]{0.49\columnwidth}
        \centering
        \includegraphics[width=\columnwidth]{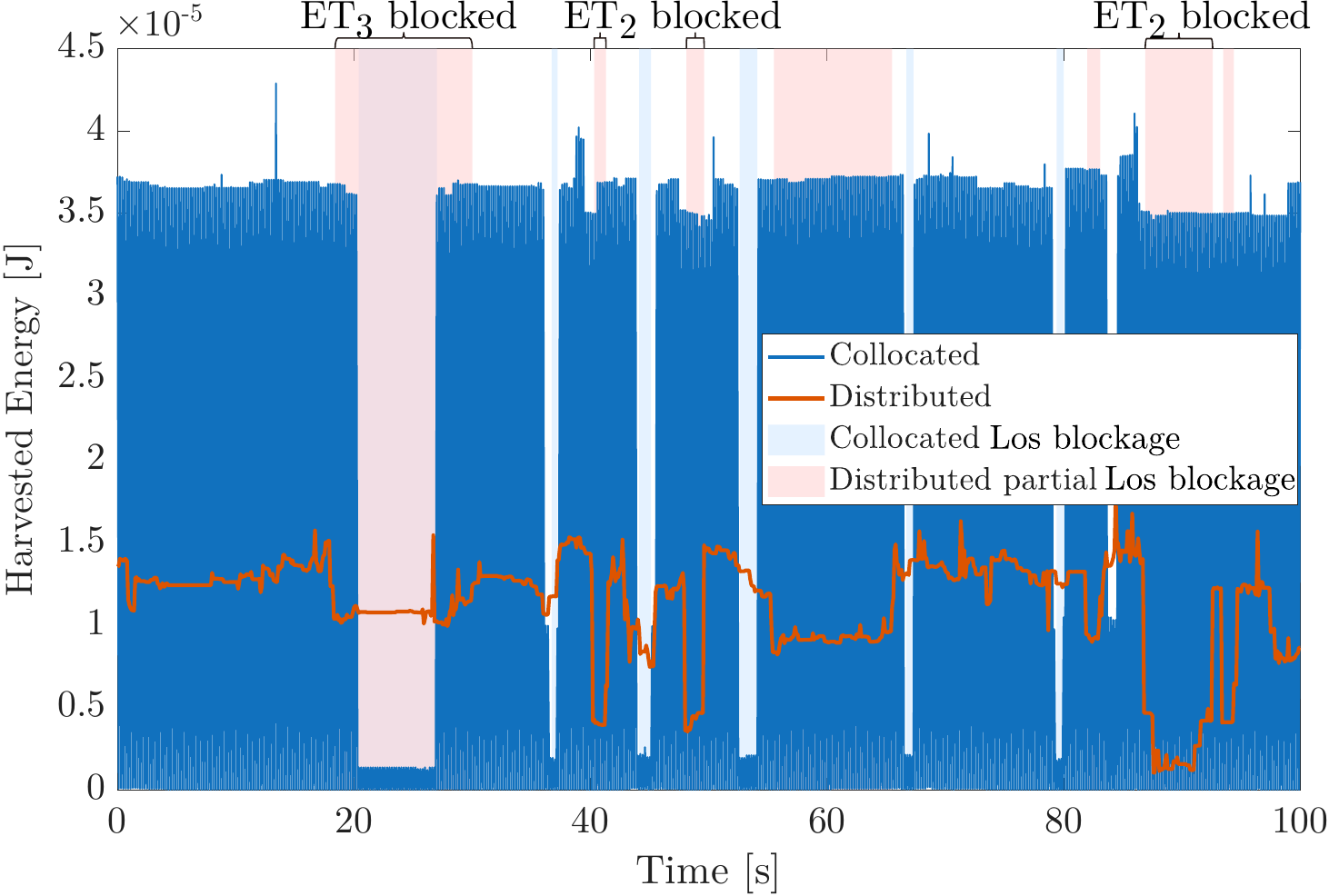}
        \caption{ER$_1$.}
        \label{subfig:ER1_scenario1}
    \end{subfigure}
    \begin{subfigure}[t]{0.49\columnwidth}
        \centering
        \includegraphics[width=\columnwidth]{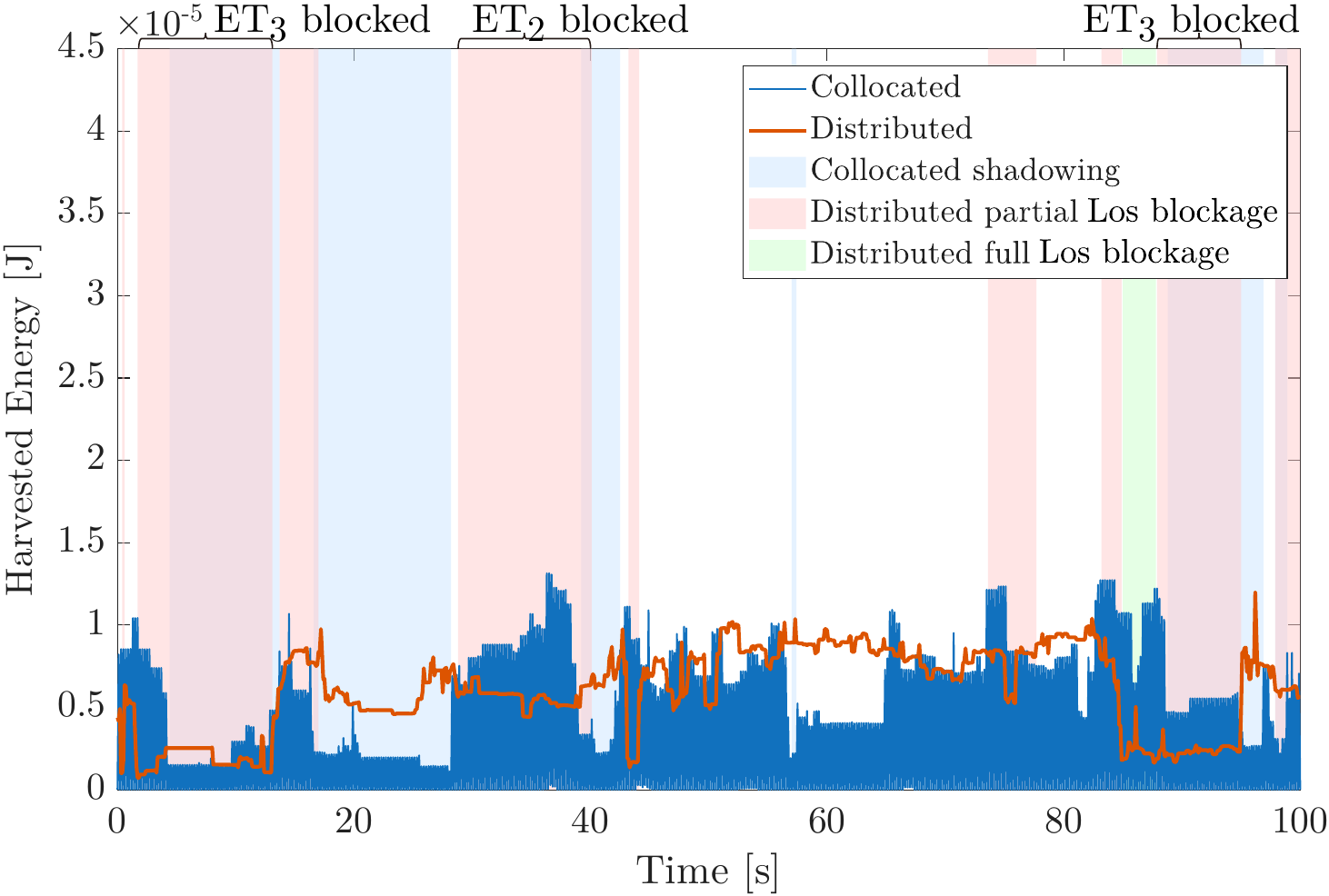}
        \caption{ER3.}
        \label{subfig:ER3_scenario1}
    \end{subfigure}
    \caption{Harvested energy per slot in the collocated and distributed WPT systems.}
    \label{fig:scenario1}
\end{figure}

Fig.~\ref{fig:scenario1} shows the time variation of the harvested energy for ER$_1$ and ER$_3$. The blue and orange curves represent the collocated and distributed systems, respectively. 
The blue shaded intervals indicate blockage of the LoS path from the collocated ET to the considered ER. The red shaded intervals indicate partial LoS blockage in the distributed system, where one of the two direct ET–ER paths is blocked. The green shaded intervals indicate full LoS blockage, where both direct paths are blocked. Reflected and other non-LoS components remain included during these intervals. 
From the figures, the harvested energy decreases during LoS blockage in both systems. 
The collocated system shows periodic slot-to-slot fluctuations due to alternating transmission and sustained energy reductions when the direct path to the considered ER is blocked. In contrast, the distributed system is less sensitive to blockage of a single direct path because the other ET can contribute through an unblocked path and multipath components. However, energy degradation is larger when the associated ET–ER path is blocked and becomes substantial when both direct paths are blocked. Thus, robustness depends on the ET–ER geometry and spatial correlation of blockage events. Optimal ET placement and blockage-aware clustering will be investigated in future work.

\begin{figure}
    \begin{subfigure}[t]{0.49\columnwidth}
        \centering
        \includegraphics[width=\linewidth]{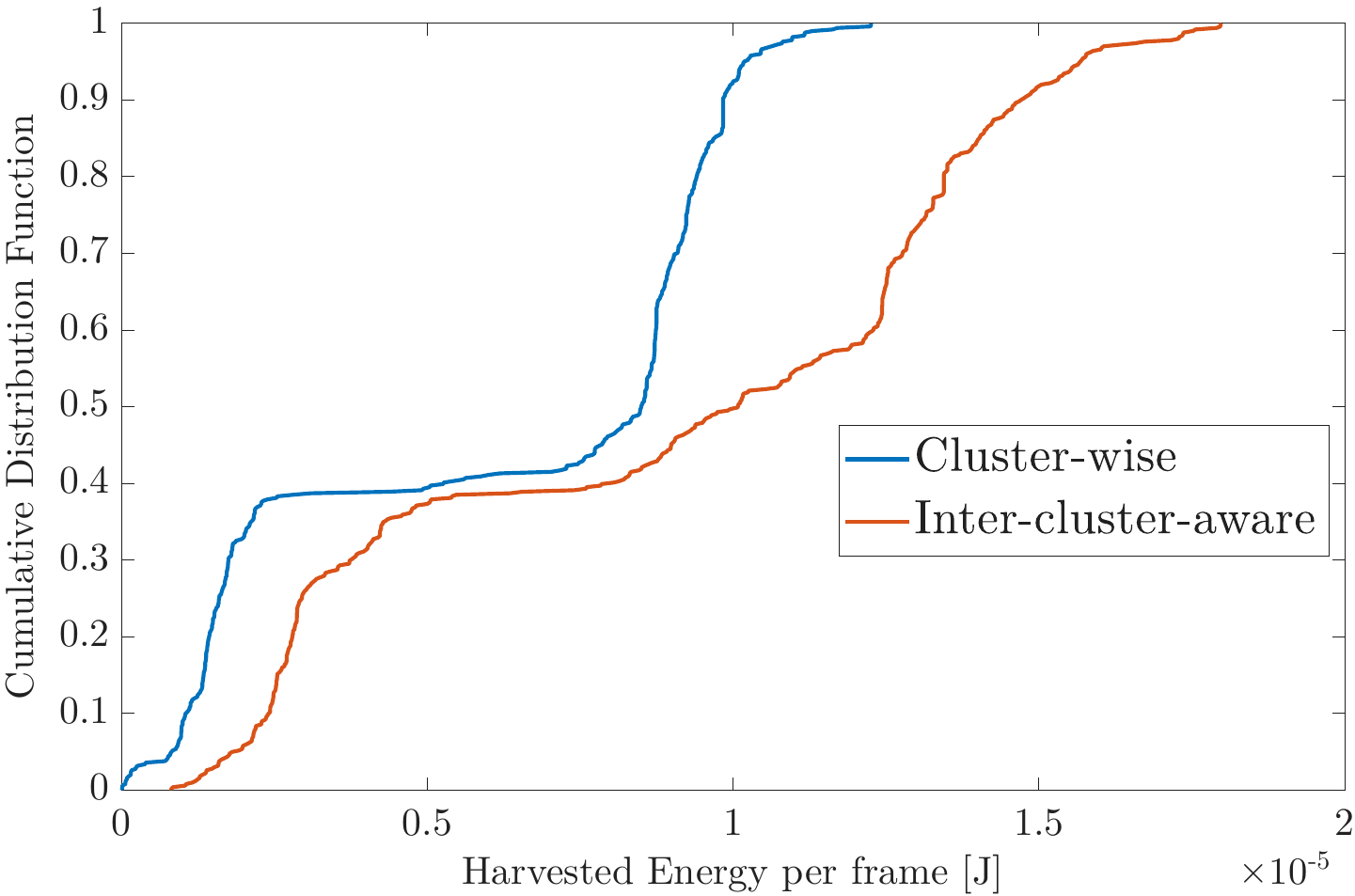}
        \caption{ER$_1$}
    \end{subfigure}
    \begin{subfigure}[t]{0.49\columnwidth}
        \centering
        \includegraphics[width=\linewidth]{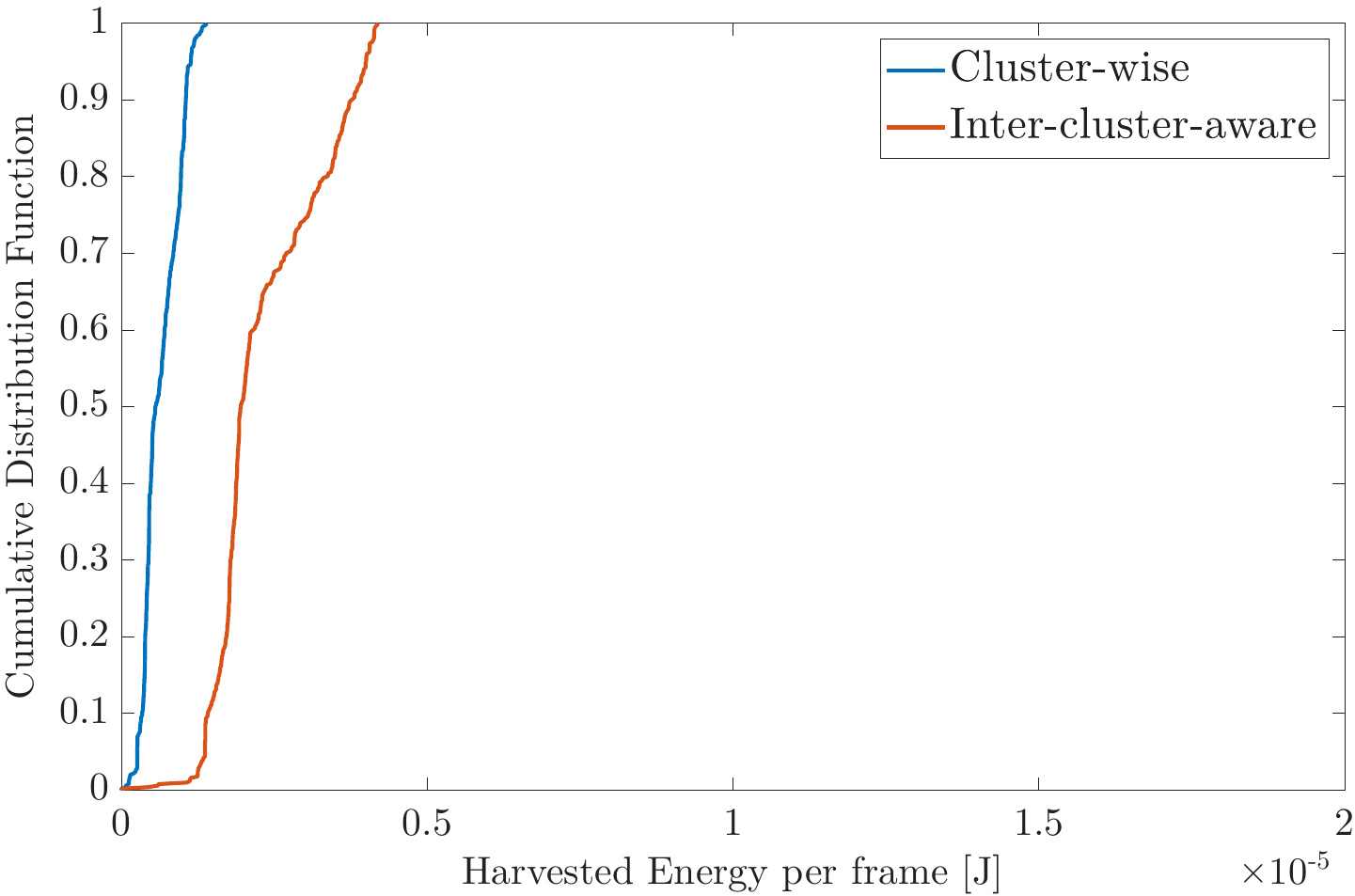}
        \caption{ER$_2$}
    \end{subfigure}
    \begin{subfigure}[t]{0.49\columnwidth}
       \centering
        \includegraphics[width=\linewidth]{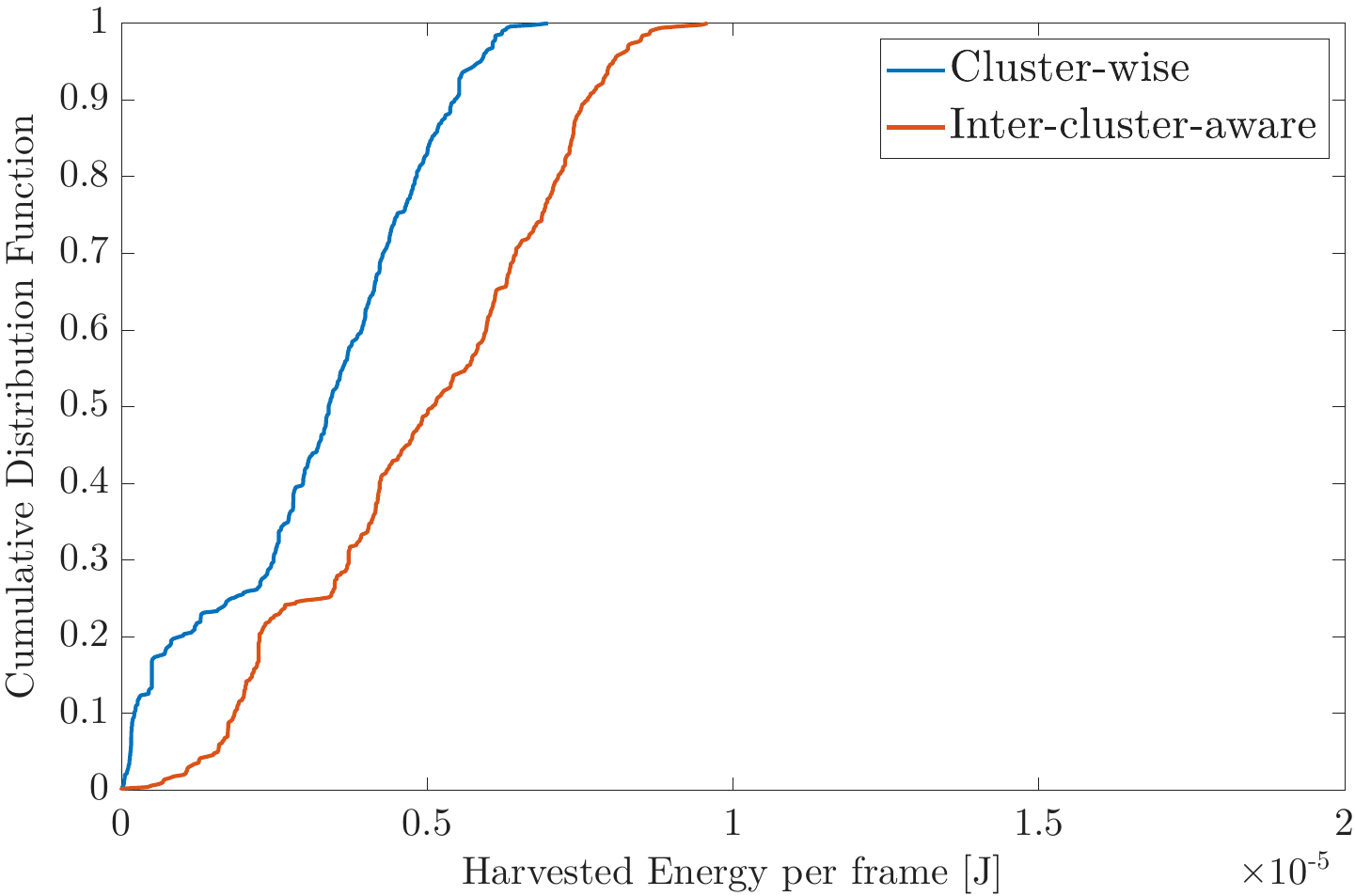}
       \caption{ER$_3$}
    \end{subfigure}
   \begin{subfigure}[t]{0.49\columnwidth}
        \centering
       \includegraphics[width=\linewidth]{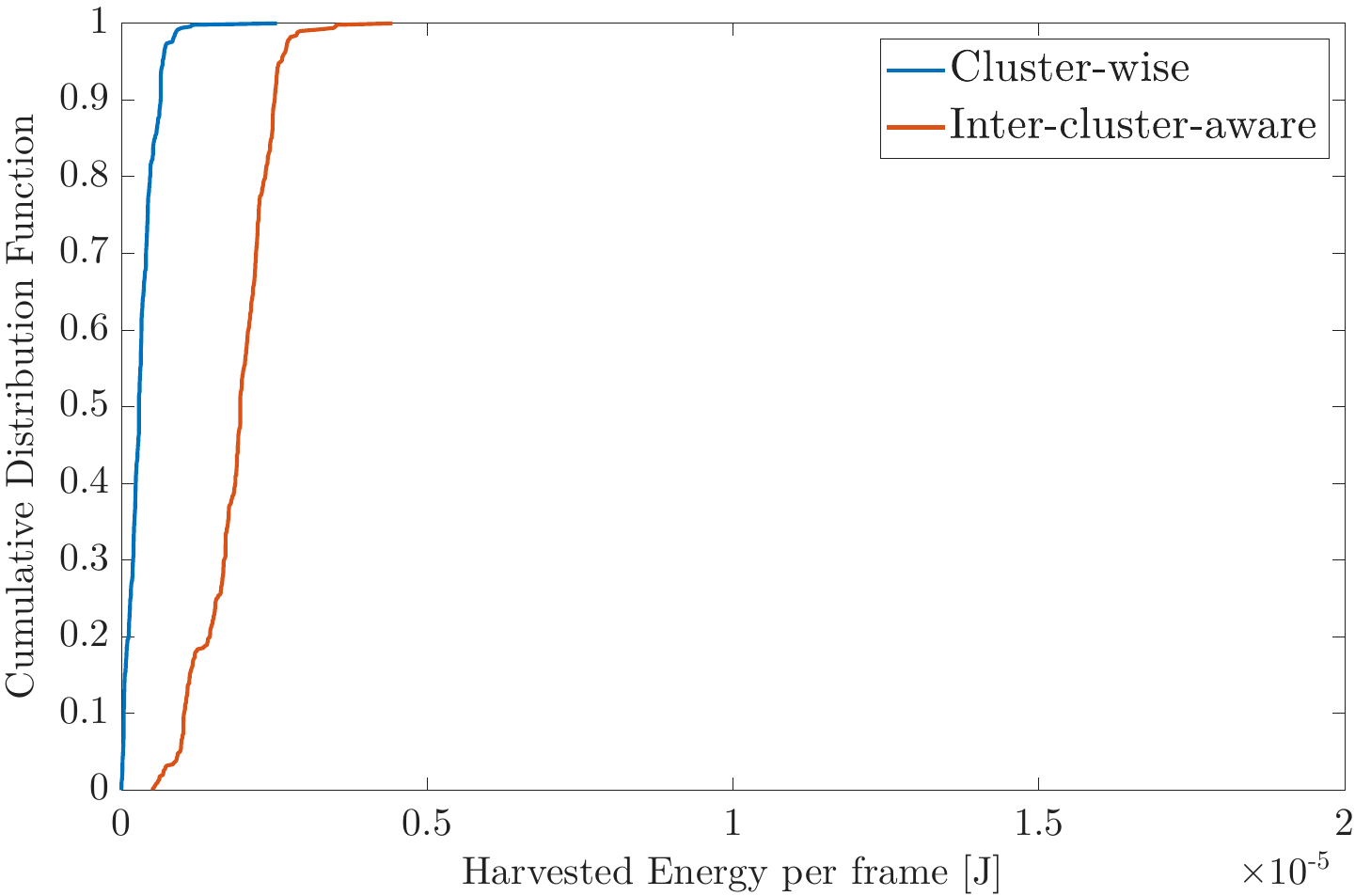}
        \caption{ER$_4$}
        \label{subfig:ER4}
    \end{subfigure}
    \caption{Cumulative distribution function~(CDF) of harvested energy per frame for the cluster-wise and inter-cluster-aware methods.}
    \label{fig:cdf}
\end{figure}

{\black 
\subsubsection{Effect of Weight Vectors Optimization}
\label{subsec:senario2}
Fig.~\ref{fig:cdf} shows the cumulative distribution function~(CDF) of harvested energy per frame in the second scenario. From the figure, the CDF of the inter-cluster-aware method is shifted toward higher harvested-energy values relative to that of the cluster-wise method. This result indicates that jointly optimizing the ET weight vectors while accounting for cross-cluster signal contributions improves the harvested energy.
The coordinated beamforming design can therefore exploit inter-cluster interference beneficially, rather than allowing the ETs to optimize their beams independently.

}

\section{Conclusion}
\label{sec:conclusion}
In this study, we proposed a cluster-based spatio-temporal scheduling framework for coordinated multi-transmitter WPT. The present simulations provided an initial evaluation of two components of the framework. First, distributed ET deployment provided alternative propagation paths under LoS blockage; however, the resulting energy degradation depended strongly on whether the direct path from the ET associated with the considered ER was blocked. This result highlights the importance of ET placement and ER clustering. Second, inter-cluster-aware beamforming shifted the harvested-energy CDFs toward higher values than independent cluster-wise MRT. The time-slot assignment component was formulated but was not separately evaluated in this study. Future work will evaluate optimized time-slot assignment, develop scalable algorithms for beamforming and scheduling.


\begin{thebibliography}{99}

    \bibitem{Clerckx2021}
    B. Clerckx, K. Huang, L. R. Varshney, S. Ulukus and M. -S. Alouini, ``Wireless Power Transfer for Future Networks: Signal Processing, Machine Learning, Computing, and Sensing,'' \emph{IEEE Journal of Selected Topics in Signal Processing}, vol. 15, no. 5, pp. 1060-1094, Aug. 2021.

    \bibitem{Huang2019} J. Huang, Y. Zhou, Z. Ning, and H. Gharavi, ``Wireless Power Transfer and Energy Harvesting: Current Status and Future Prospects,'' \emph{IEEE Wireless Communications}, vol.~26, no.~4, pp.~163--169, Aug.~2019.

    \bibitem{Shen2021} S. Shen and B. Clerckx, ``Joint Waveform and Beamforming Optimization for MIMO Wireless Power Transfer,'' \emph{IEEE Transactions on Communications}, vol.~69, no.~8, pp.~5441--5455, Aug.~2021.

    \bibitem{Zhou2018}
    S. Zhou, X. Wang, N. Cao and X. Li, ``Performance Analysis of Wireless Powered Communications With Multiple Antennas,'' \emph{IEEE Access}, vol. 6, pp. 15541-15549, 2018.
    
    \bibitem{Bayat2022}
    A. Bayat and S. A\"{i}ssa, ``Fair Scheduling of Wireless Power Transfer to Nonlinear Energy Harvesters,'' \emph{IEEE Transactions on Green Communications and Networking}, vol.~6, no.~2, pp.~1096--1106, Jun. 2022.

    \bibitem{Sawada2024}
    Y. Sawada, S. Shiraki, T. Matsuda, T. Hiraguri, K. Maruta and T. Kimura, ``Time Division Wireless Power Transfer Using Receiver Grouping,'' \emph{IEEE Access}, vol.~12, pp.~109930--109942, Aug.~2024.

    \bibitem{Sawada2025}
    Y. Sawada, S. Shiraki, T. Matsuda, T. Hiraguri, K. Maruta and T. Kimura, ``Battery-Aware Time Division Wireless Power Transfer Based on Channel State,'' \emph{2025 Sixteenth International Conference on Ubiquitous and Future Networks (ICUFN)}, Lisbon, Portugal, 2025, pp. 417--423.

    \bibitem{Huang2019-2}
    Y. Huang, Y. Liu and G. Y. Li, ``Energy Efficiency of Distributed Antenna Systems With Wireless Power Transfer,'' \emph{IEEE Journal on Selected Areas in Communications}, vol.~37, no.~1, pp.~89--99, Jan. 2019.

    \bibitem{Kim2019}
    K.-W. Kim, H.-S. Lee, and J.-W. Lee, ``Waveform Design for Fair Wireless Power Transfer With Multiple Energy Harvesting Devices,'' \emph{IEEE Journal on Selected Areas in Communications}, vol.~37, no.~1, pp.~34--47, Jan.~2019.
    
    \bibitem{Lambert}
    R. M. Corless, G. H. Gonnet, D. E. Hare, D. J. Jeffrey, and D. E. Knuth,
    ``On the Lambert W function,'' \emph{Advances in Computational Mathematics},
    vol.~5, no.~1, pp.~329--359, Dec. 1996. 

    \bibitem{Chen2023}
    V. X. Chen and J. N. Hooker, ``A Guide to Formulating Fairness in an Optimization Model,'' 
    \emph{Annals of Operations Research}, vol.~326, pp.~581--619, Apr.~2023.

    \bibitem{SionnaRT}
    J. Hoydis et al., ``Sionna RT: Differentiable Ray Tracing for Radio Propagation Modeling,'' \emph{Proc. 2023 IEEE GLOBECOM Workshops (GC Wkshps)}, pp.~317--321, Dec.~2023.


\end{thebibliography}
\end{document}